# BIO-CXRNET: A Robust Multimodal Stacking Machine Learning Technique for Mortality Risk Prediction of COVID-19 Patients using Chest X-Ray Images and Clinical Data

Tawsifur Rahman[1], Muhammad E. H. Chowdhury[1*], Amith Khandakar[1], Zaid Bin Mahbub[2*], Md Sakib Abrar Hossain[3], Abraham Alhatou[4], Eynas Abdalla[5], Sreekumar Muthiyal[6], Khandaker Farzana Islam[1], Saad Bin Abul Kashem[7], Muhammad Salman Khan[1], Susu M. Zughaier[8], Maqsud Hossain[3]

[1]Department of Electrical Engineering, Qatar University, Doha, Qatar, P.O. Box 2713
[2]Department of Physics and Mathematics, North South University, Dhaka-1229, Bangladesh
[3]NSU Genome Research Institute (NGRI), North South University, Dhaka-1229, Bangladesh
[4]Department of Biology, University of South Carolina (USC), Columbia, SC 29208, United States
[5]Anesthesia Department, Hamad General Hospital, P.O. Box 3050
[6]Department of Radiology, Hamad General Hospital, P.O. Box 3050
[7]Department of Computer Science, AFG College with the University of Aberdeen, Doha, Qatar
[8]Department of Basic Medical Sciences, College of Medicine, QU Health, Qatar University, Doha, Qatar, PO Box 2713
*Corresponding author: Muhammad E. H. Chowdhury (mchowdhury@qu.edu.qa), Zaid Bin Mahbub (zaid.mahbub@northsouth.edu)

## Abstract

Fast and accurate detection of the disease can significantly help in reducing the strain on the healthcare facility of any country to reduce the mortality during any pandemic. COVID-19 detection is based on physical examinations, radiological imaging, blood testing, and reverse transcription polymerase chain reaction (RT-PCR) method. Severity of COVID-19 has been detected using clinical data, radiological imaging and sign-symptoms. The goal of this work is to create a multimodal system using a novel machine learning framework that uses both Chest X-ray (CXR) images and clinical data to predict severity in COVID-19 patients. In addition, the study presents a nomogram-based scoring technique for predicting the likelihood of death in high-risk patients. This study uses 25 biomarkers and CXR images in predicting the risk in 930 COVID-19 patients admitted during the first wave of COVID-19 (March-June 2020) in Italy. The proposed multimodal stacking technique produced the precision, sensitivity, and F1-score, of 89.03%, 90.44%, and 89.03%, respectively to identify low or high-risk patients. This multimodal approach improved the accuracy by 6% in comparison to the CXR image or clinical data alone. Finally, nomogram scoring system using multivariate logistic regression- was used to stratify the mortality risk among the high-risk patients identified in the first stage. Lactate Dehydrogenase (LDH), $O_2$ percentage, White Blood Cells (WBC) Count, Age, and C - reactive protein (CRP) were identified as useful predictor using random forest feature selection model. Five predictors parameters and a CXR image based nomogram score was developed for quantifying the probability of death and categorizing them into two risk groups: survived (<50%), and death (>=50%), respectively. The multi-modal technique was able to predict the death probability of high-risk patients with an F1 score of 92.88 %. The area under the curves for the development and validation cohorts are 0.981 and 0.939, respectively. The recommended clinical parameter and CXR-based prognostic model, which is also available in the form of a website, can aid medical doctors in improving the management of COVID-19 patients by predicting mortality risk early.

## I. Introduction

As of May 6, 2022, over 6.2 million individuals have died and over 516 million people infected due to the COVID-19 pandemic [1]. The global corporate, economic, and social dynamics were all affected. Flight limitations, social isolation, and increased hygiene awareness have been implemented by governments all over the globe. COVID-19 is sometimes can be confused with other viral infections [2, 3], making the identification difficult. Reverse-transcription polymerase chain reaction (RT-PCR) arrays is the approved primary diagnostic approach for COVID-19 detection [4, 5]. Its detection performance can suffer due to sample contamination/damage or viral alterations in the COVID-19 genome [6, 7]. As a result, some studies [8, 9] have suggested that Chest Computed Tomography (CT) imaging can be used as an alternative approach. Besides, for a RT-PCR negative patient with COVID-19 symptoms, several researchers have recommended to use CT scan as a follow-up test [8-10]. CT scans, despite their superior performance, have several disadvantages and limitations. For early COVID-19 instances, their sensitivity is limited, image collection is slow and expensive. Chest X-ray (CXR) imaging, in contrast to CT, is a less expensive, faster, and more widely available technique that exposes the body to less hazardous radiation. [11]. Chest X-rays are frequently used as an alternative COVID-19 screening technique, and it has been demonstrated to have a high predictive value [12]. On radiological images, early COVID-19 cases showed bilateral, multifocal ground-glass opacities (GGO) with posterior or peripheral distribution, primarily in the lower lung lobes, and eventually progressed into pulmonary consolidation [13, 14]. The lung abnormalities share many common characteristics. As a result, doctors have a hard time distinguishing between COVID-19 infection and other types of viral pneumonia. As a result, in the current scenario, this symptom likeness could result in a misdiagnosis, delayed treatment, or can even result in death.

In recent years, great advances in Deep Learning approaches have resulted in state-of-the-art performance in a variety of Computer Vision applications, including picture classification, object recognition, and image segmentation. As a result of this achievement, deep learning-based solutions have become more widely used in a variety of disciplines. With the advent of deep Convolutional Neural Networks (CNNs), their use on CXR images is widely researched and accepted. Rajpurkar et al. [15] proposed the CheXNet network, by modifying Densenet121 on one of the largest Chest X-ray datasets [16], containing 100 thousand X-ray images for 14 different diseases. Similarly, Rahman et al. [17] trained CXRs for detecting pulmonary tuberculosis (TB) using a dataset of 3,500 infected and 3,500 normal CXRs. They have also re-trained the DenseNet201 network on TB and normal datasets and obtained state-of-the-art performance in TB detection with a sensitivity of 98.57 %. However, until recently, lung segmentation is used as the first step in their detection technique [18, 19], which helps in localizing the decision-making area for the machine learning networks. They have used the popular dataset from the Montgomery [20] and Shenzhen [21] CXR lung mask datasets, which together produce 704 X-ray images for Normal and TB patients. However, sometimes due to severe deformity of the lungs in extreme COVID-19 cases or low resolution pictures, the segmentation performance suffers. Khuzani et al [22] proposed that a set of features of CXR pictures may be constructed using the dimensionality reduction method to build an effective machine learning classifier that can identify COVID-19 instances from non-COVID-19 cases with high accuracy and sensitivity. Mathew et al. [23] proposed a Siamese neural network-based severity score to automatically measures radiographic COVID-19 pulmonary disease severity which was verified with pulmonary x-ray severity (PXS) scores from two thoracic radiologists and one in-training radiologist. Kim et al. [24] proposed a completely automated triage pipeline that analyses chest radiographs for the presence, severity, and progression of COVID-19 pneumonia and produced an accuracy of 79.9%. Maguolo and Nanni in [25] questioned the performance of COVID -19 detection from X-rays in various literature and mentioned that it should include larger and diverse X-rays to avoid biases. Robert et al. [26] have reasoned in the same line of thought by doing an extensive literature review and suggesting the use of a diverse and large dataset for the proposal of COVID-19 detection from chest X- Rays. The authors of this paper were one of the pioneers that have proposed state-of-the-art deep learning model to detect pneumonia [27] and COVID-19 [28] from Chest X-rays. They have further improved their work in [29], where they created the largest benchmark dataset with 33,920 CXR images, including 11,956 COVID-19 samples using an effective human-machine collaborative strategy to annotate ground-truth lung segmentation masks. This is the largest CXR lung segmentation dataset to the best of the authors' knowledge, which can help in CXR-related computer-aided-diagnostic tools development using deep learning methods. In this study, the authors have used the model trained on that state-of-the-art dataset to segment the lung areas from the CXR images. We investigated the impact of image enhancement techniques on segmented lungs for COVID-19 prediction in a previous study [30], confirming that gamma correction enhancement provided an F1-score of around 90% using a dataset of total 18,479 Chest X-ray images (8851 normal, 6012 non-COVID other lung diseases, and 3616 COVID-19) and their ground truth lung masks.

According to recent research, biomarkers can play a dominating role in giving critical information about an individual's health and in identifying COVID-19. Sarah et al. [31] presented the Kuwait Progression Indicator (KPI) score as a predictive tool for estimating the severity of COVID-19 progression. In contrast to scoring systems that rely on self-reported symptoms and other subjective characteristics, the KPI model was founded on laboratory variables, which are objectively measurable metrics. Patients were classified as low risk if their KPI score goes below -7 and as high risk if it increases beyond 16, however, the risk of advancement was unknown for those with a score between -6 and 15. This restricts its applicability to a large variety of patient groups. Weng et al. [32] published an early prediction score named ANDC to predict mortality risk in COVID patients. This prediction model was developed using data from 301 adult patients with laboratory-confirmed COVID-19. LASSO regression indicated age, neutrophil-to-lymphocyte ratio, D-dimer, and C-reactive protein acquired on hospital admission as strong predictors of death for COVID-19 patients. The area under the curve (AUC) for the derivation and validation cohorts was 0.921 and 0.975, respectively, indicating that the nomogram was well calibrated and discriminated. COVID patients were categorized into three groups based on ANDC cut-off values of 59 and 101. The low-risk group (ANDC 59) had a mortality probability of less than 5%, the moderate risk group (59<ANDC<101) had a death probability of 5% to 50%, and the high-risk group (ANDC>101) had a death probability of more than 50%. Xie et al. [33] developed a predictive model that combines age, lactate dehydrogenase (LDH), lymphocyte count, and SpO2 as independent predictors of death using a dataset of 444 patients, where it showed good performance for both internal (c=0.89) and external (c=0.98) validations. However, the model showed over-prediction in low-risk persons while under-prediction for the patients with high-risk.

Intensive care units (ICUs) are crucial to save severe COVID-19 patients by providing oxygen, 24-hour monitoring, care, and assisted ventilation when necessary. As a result, in areas where COVID-19 infection rate is high, ICU beds are a valuable resource [34-36]. Basic blood tests and vital signs measurements are among the routinely gathered healthcare data, which are often available within the first hour of visit to the hospitals. These data provide the patterns of changes in COVID-19 patients as described in several retrospective observational studies [37-39]. These studies have concluded that variables including alanine aminotransferase (ALT), lymphocyte count, D-dimer, C-reactive protein (CRP), and bilirubin concentrations as important clinical indicators. Therefore, clinical biomarkers can be used for developing very good prognostic model using classical and deep learning models.

Although Convolutional Neural Networks (CNNs) can be trained to stratify different diseases using radiographic or other images, imaging data alone cannot consistently identify the underlying medical cause. A combination of patient symptoms, physical exam findings, laboratory data, and radiologic imaging findings can be used to determine the underlying etiology and severity (if available). As a result, machine learning algorithms that combine information from Chest X-rays with additional clinical data from the electronic health record (EHR) will be able to better predict the severity of the patient. However, attempts to combine EHR data and imaging data for machine learning applications in healthcare have not been well-studied. Few studies have used both radiographic images and clinical biomarkers data to predict the prognostication of COVID-19 patients using artificial intelligence. Jiao et al. [40] developed a machine learning model using clinical data and CXR images to envisage COVID-19 severity and progression and reported an AUC of 82 %. Chieregato et al. [41] proposed a multimodal approach based on CT images and clinical parameters, which were supplied to Boruta feature selection algorithm with SHAP (SHapley Additive exPlanations) values, and then CatBoost gradient boosting classifier showed an AUC of 0.949 on the holdout test set for reduced features. With a probability score based on SHAP feature importance, the model intends to give clinical decision support to medical physicians. However, the reported studies either produced poor performance or have used small datasets, which limits the generalizability of the models, or have used CT modality, which has limitations.

The above pitfalls have encouraged the authors of this work to develop a multimodal system using CXR and clinical biomarker based system to stratify the severity of COVID-19 patient and their risk of death. This is one of the first works, which used both CXR and biomarkers to develop a COVID-19 severity prediction model. Thus, the work tackles the above limitation with the below contributions:
- We created a machine learning model that used chest radiographs and clinical data to identify severe infection and death risks in COVID-19 patients.
- Combining multimodal data (image and clinical data), it was possible to boost the performance of the prognostic model significantly compared to the model using individual data.
- A nomogram-based scoring system to combine image features and important biomarkers along with a novel stacking machine learning technique is proposed.
- A web application is developed using the proposed multimodal approach, which will enable the clinicians to utilize such a tool to aid in the diagnostic process.

The rest of the article has the following structure: Section II describes the methodology of the study, which includes dataset description, preprocessing stages, machine learning and stacking technique, and nomogram-based scoring system development. Section III presents the findings of the experiments and reports the performance of the scoring technique, while Section IV explains the findings. Finally, Section V concludes the article with future recommendations.

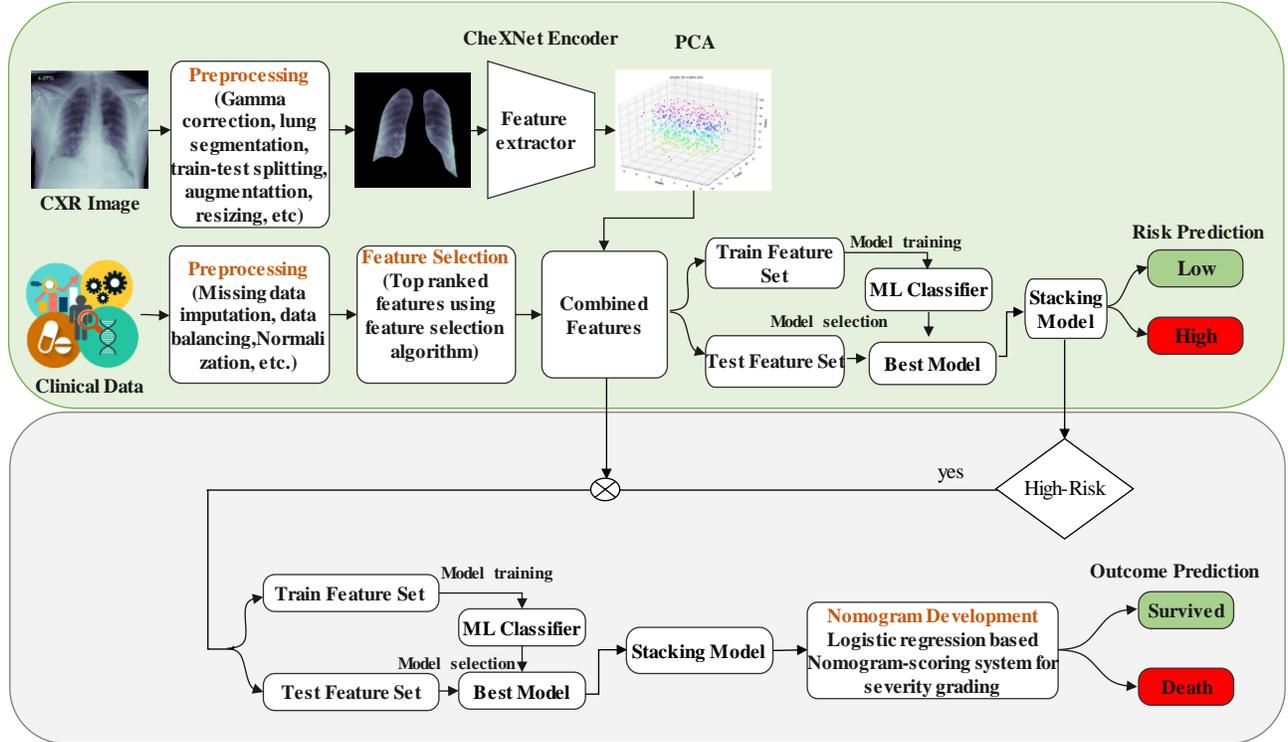

**Figure 1:** Overview of the methodology.

## II. Methodology

Two major investigations were carried out in this study. In the first investigation, a multimodal approach using CXR image and clinical data was used to predict the severity risk of COVID-19 patients. Firstly, CXR images are preprocessed and the lung area is segmented and then fed to a pre-trained deep CNN model to extract image features, and then principal component analysis (PCA) was used to reduce the dimensionality of the extracted image features. In parallel, clinical data was processed and the clinical features were ranked using a feature selection algorithm. Finally, the PCA components and top-ranked clinical features were combined to develop a stacking ensemble model to predict the low or high-risk patients. In the second investigation, we analyzed high-risk patients' data to predict the death outcome using the stacking model from CXR images and clinical biomarkers. Moreover, we established a scoring technique using nomogram for the early prediction of death outcomes. Figure 1 illustrates the schematic overview of the methodology.

**Dataset Description**
The study utilized a dataset from the first wave of COVID-19, between March and June 2020, that included both CXRs and clinical data obtained from six Italian hospitals at the time of admission for symptomatic COVID-19 patients [42]. This dataset contains an anteroposterior (AP) or posteroanterior (PA) view of 930 X-ray images and clinical data of COVID-19 positive patients [42]. Each of the patients was tested RT-PCR positive for COVID-19. This dataset consists of 396 (42.6%) low and 534 (57.4%) high-risk patients. Moreover, 364 (68.2%) out of 534 high-risk patients survived while 170 (31.8%) high-risk patients died. Figure 2 illustrates sample CXR images from the dataset.

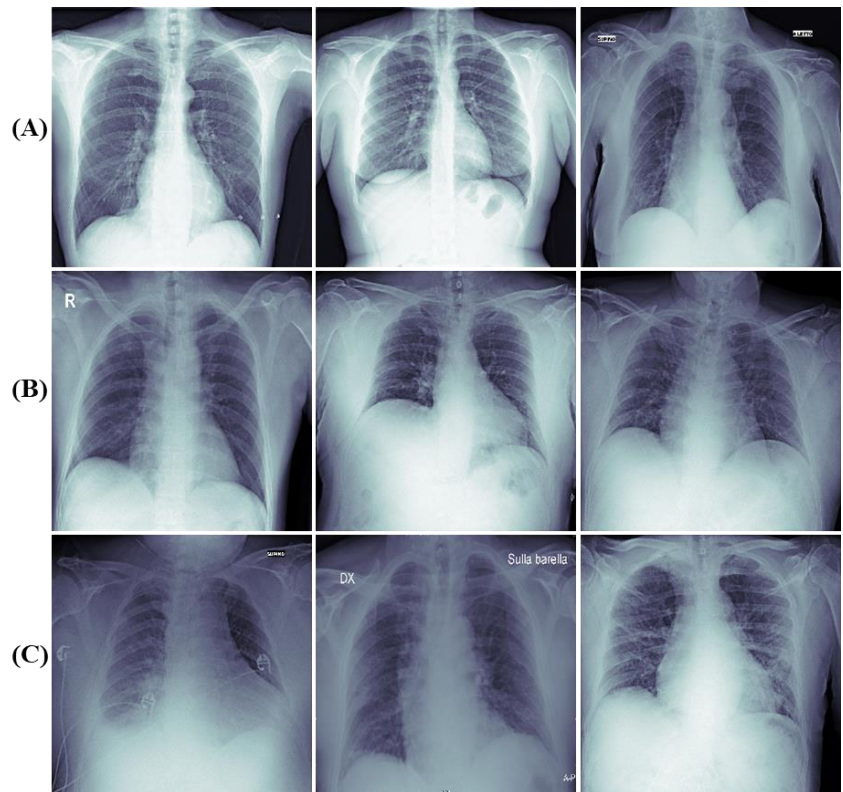

**Figure 2:** Chest X-ray sample images for COVID-19 (A) Low-risk patients, (B) High-risk patients with survival outcomes, and (C) High-risk patients with death outcomes.

**Statistical Characteristics**

Stata/MP 13.0 software was used to conduct a statistical analysis of the patient's demographic, signs and symptoms, clinical data, comorbidity, and outcome. Gender, age, and twenty-three signs and symptoms, comorbidity, and clinical biomarkers are available in the dataset. Table 1 lists the statistical features of 25 parameters (age, gender, sign and symptoms, comorbidity, clinical biomarkers). Gender is represented in numbers and percentages. The number of missing data (N), presence and absence of signs and symptoms, mean (M), and standard deviation (SD) were reported for the remaining variables. Univariate analysis (Chi-square test) was done for gender, while the rest of the variables were subjected to Wilcoxon's ranked tests. The p-value was considered significant if it is less than 0.05 using a 95% statistical significance criterion.

**Table 1:** Summary of statistical characteristic of the study patients

| Item | Low-risk | High-risk | Total | Statistics | P-value |
|---|---|---|---|---|---|
| **Gender** | | | | $\chi^2=16.45$ | <.05 |
| • Male (%) | 159 (40%) | -149 (28%) | 308 (33%) | | |
| • Female (%) | 237 (60%) | 385 (72%) | 622 (67%) | | |
| **Age (years)** | | | | Z=-8.5 | <.05 |
| • N | 396 (0) | 534(0) | 930 (0) | | |
| • M ± SD | 60.3±15.9 | 67.8±13.5 | 64.7±14.9 | | |
| **Body temperature (°C)** | | | | Z=-7.55 | <.05 |
| • N | 354 (42) | 456 (78) | 810 (120) | | |
| • M ± SD | 37.5±0.93 | 39.6±1.3 | 37.56±0.98 | | |
| **Cough (%yes)** | | | | Z=-6.67 | <.05 |
| • N | 395 (1) | 530 (4) | 925 (5) | | |
| • Yes/no | 210/185 | 460/70 | 670/255 | | |
| **Difficulty in Breathing** | | | | Z=-9.35 | <.05 |
| • N | 395 (1) | 531(3) | 926 (4) | | |
| • Yes/No | 180/215 | 450/81 | 630/296 | | |
| **Red Blood Cell (10^9L)** | | | | Z=-16.56 | <.05 |
| • N | 371 (25) | 515 (19) | 886 (44) | | |
| • M ± SD | 4.68±0.7 | 4.56±0.71 | 4.59±0.7 | | |
| **White blood cell count (10^9 L)** | | | | Z=-8.77 | <.05 |
| • N | 386 (10) | 524 (10) | 910 (20) | | |
| • M ± SD | 6.08±2.76 | 7.87±4.36 | 7.1±3.87 | | |
| **CRP (mg/dL)** | | | | Z=-3.53 | <.05 |
| • N | 377 (19) | 514 (20) | 891(39) | | |
| • M ± SD | 23.01±43.3 | 39.5±69.4 | 32.5±60.33 | | |
| **Fibrinogen (mg/dL)** | | | | Z=0.174 | 0.912 |
| • N | 75 (321) | 141 (393) | 216 (714) | | |
| • M ± SD | 561.07±115 | 641.19±172 | 613.37±159 | | |
| **Glucose (mg/dL)** | | | | Z=-11.84 | <.05 |
| • N | 302 (94) | 439 (95) | 741(189) | | |
| • M ± SD | 114.5±48.2 | 130.9±60.32 | 124.26±56 | | |
| **LDH (U/L)** | | | | Z=-3.28 | <.05 |
| • N | 291(105) | 404 (130) | 695 (235) | | |
| • M ± SD | 282.4±114 | 442.4±272 | 375.4±234 | | |
| **INR** | | | | Z=-7.93 | <.05 |
| • N | 255 (141) | 413 (121) | 668 (262) | | |
| • M ± SD | 1.15±0.29 | 1.3±0.76 | 1.24±0.63 | | |
| **D-dimer** | | | | Z=-1.81 | 0.064 |
| • N | 106 (290) | 144 (390) | 250 (680) | | |
| • M ± SD | 1055.8±1384 | 3780.6±8635 | 2625.3±6741 | | |
| **$O_2$ Percentage (%)** | | | | Z=-6.42 | <.05 |
| • N | 289 (107) | 365 (169) | 654 (276) | | |
| • M ± SD | 95.4±3.72 | 89.5±8.01 | 92.3±7.02 | | |
| **$PaO_2$ (mmHg)** | | | | Z=-8.33 | <.05 |
| • N | 288 (108) | 412 (122) | 800 (130) | | |
| • M ± SD | 75.5±17.15 | 69.95±31.26 | 72.21±26.5 | | |
| **$SaO_2$ (%)** | | | | Z=1.36 | 0.845 |
| • N | 165 (231) | 212(322) | 377 (553) | | |
| • M ± SD | 94.92±4.38 | 89.95±8.92 | 92.13±7.69 | | |
| **$PaCO_2$ (mmHg)** | | | | Z=5.36 | <.05 |
| • N | 278 (118) | 403 (131) | 681 (249) | | |
| • M ± SD | 33.49±5.4 | 33.1±6.9 | 33.26±6.34 | | |
| **pH** | | | | Z=-2.01 | <.05 |
| • N | 271 (125) | 386 (148) | 657 (273) | | |
| • M ± SD | 7.45±0.05 | 7.15±0.06 | 7.35±0.05 | | |
| **Cardiovascular Disease** | | | | Z=7.78 | <.05 |
| • N | | | | | |
| • Yes/No | 335 (61) | 467 (67) | 802 (128) | | |
| | 230/105 | 345/122 | 575/227 | | |
| **Heart Failure (%)** | | | | Z=6.40 | <.05 |
| • N | 333 (63) | 465 (69) | 798 (132) | | |
| • Yes/No | 215/118 | 288/187 | 503/305 | | |
| **High Blood Pressure** | | | | Z=-5.66 | <.05 |
| • N | 337(59) | 467 (67) | 804 (126) | | |
| • Yes/No | 227/110 | 330/337 | 557/447 | | |
| **Cancer** | | | | Z=-1.84 | 0.067 |
| • N | 337 (59) | 467 (67) | 804 (126) | | |
| • Yes/No | 112/225 | 164/303 | 276/528 | | |
| **Chronic Kidney Disease (%)** | | | | Z=-5.81 | <.05 |
| • N | 337 (59) | 467 (67) | 804 (126) | | |
| • Yes/No | 156/181 | 320/147 | 476/328 | | |
| **Respiratory Disease** | | | | Z=0.177 | .865 |
| • N | 262 (134) | 290(244) | 552 (378) | | |
| • Yes/No | 162/100 | 195/95 | 357/195 | | |

## Data Preprocessing

This section discusses the data preprocessing steps for both of the data modalities in detail.

*Chest X-ray Image Preprocessing*

*A. Gamma Correction*

Image enhancement is a common picture-processing technique that emphasizes significant information in an image while reducing or removing other information to increase the quality of identification. Gamma correction was applied to CXRs as shown in our previous work [30] that it enhances COVID detection performance by improving image quality. Linear operations, such as pixel-wise scalar multiplication, addition, and subtraction are often performed for image normalization, whereas Gamma correction technique is a non-linear operation performed on the pixels of the source image. Gamma correction uses a projection link with gamma value and pixel value as per the internal map. Here pixel value can vary between 0 to 255. If G is the grey scale value, then the output pixel after gamma correction s(G) can be written as:

$$s(G) = 255 \left(\frac{G}{255}\right)^{1/\gamma(G)} \quad (1)$$

where $\gamma(G)$ represents gamma value.

*B. Lung Segmentation*

As discussed earlier, it is very important to localize the region of interest for the machine learning networks, i.e. the lungs in the Chest X-rays in this case. Feature Pyramid Networks (FPN) [43] segmentation network with DenseNet121 [44] encoder as a backbone outperformed other conventional segmentation networks in our previous work for CXR lung segmentation [29]. In [29], a detailed investigation was done on three segmentation architectures, U-Net [45], U-Net++ [46], and Feature Pyramid Networks (FPN) [43] with different encoder backbones. Using the FPN network with DenseNet121 backbone, it segmented the lung area very accurately which was verified by the experienced radiologists as well. Some of the X-ray images and their corresponding masks are illustrated in Figure 3.

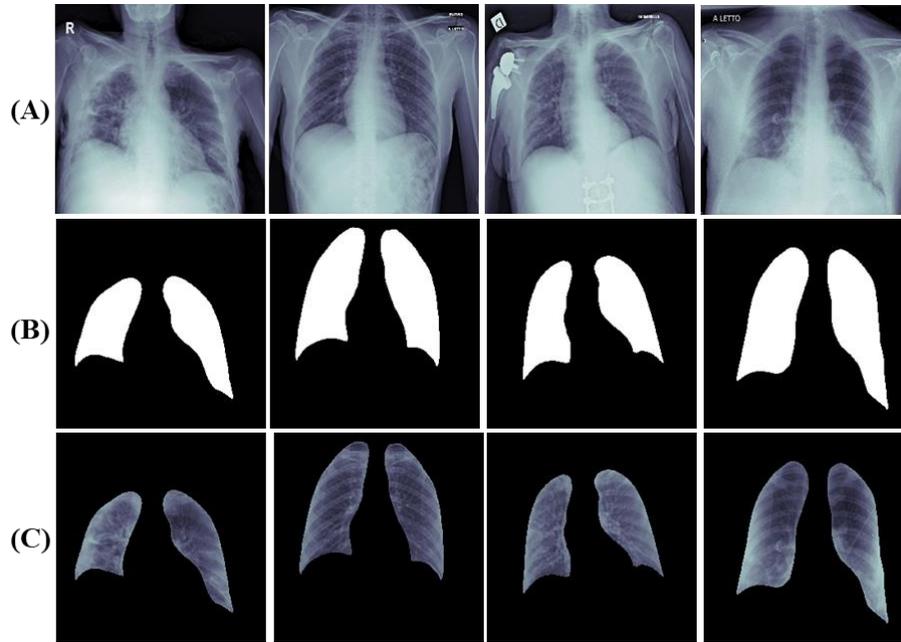

**Figure 3:** Samples X-ray images from the study dataset (A), generated masks by the best performing densenet121 FPN model (B) and corresponding segmented lung (C).

*C. Feature Extraction*
A ChexNet CNN model which is based on DenseNet-121 [44] architecture was used to extract important features from the segmented Chest X-rays. It should be worth mentioning here that CheXNet is a variant of DenseNet which was trained on a large Chest X-ray dataset and the pretrained model is available publicly. It performed exceptionally well-COVID-19 classification task as shown in our previous work [30]. To extract useful features from the segmented lung area of the CXR images, features from the last layer ('AvgPool') before the Softmax layer of the CheXNet model were extracted.

*D. PCA for Dimensionality Reduction*
A feature reduction technique called Principal Component Analysis (PCA) was applied to reduce the dimensionality of the feature space produced from the ChexNet model. It projects high-dimensional data into a new lower-dimensional representation with as minimal reconstruction error as possible. Because all the fundamental components in the reduced set are orthogonal to one another, there is no redundant data. PCA was calculated with whitening, which can improve accuracy by forcing data to meet certain assumptions.

*Clinical Data Preprocessing*
*A. Data Imputation and Normalization*
The most critical phase in clinical data preprocessing for machine learning model construction is missing data imputation. Many blood biomarkers were obtained for each patient, and many of them were absent from certain patients. Instead of eliminating the missing data for the various variables, different imputation techniques were investigated. Deleting the missing variable can result in the loss of critical and contextual information, as well as affecting the dataset's generalized representation [20]. Machine learning (ML)-based data imputation methods have become more popular for missing value imputation. On the other hand, this technique necessitates the creation of a distinct model for each missing data column. A popular data imputation technique, called multivariate imputation by chained equations (MICE) was used in this study for dealing with missing data. It is reported in literature and authors' previous works [47-50] that the MICE technique outperforms other imputation techniques for clinical data [51].

The effectiveness of machine learning models for generalized performance is strongly reliant on the quality of the input data. The term "data normalization" refers to the process of scaling or changing data so that each feature contributes equally to the training process. Normalization has been demonstrated to increase the performance of machine learning models in numerous studies [29]. In this investigation, Z-score normalization was used by subtracting the average of data and dividing it by standard deviation.

*B. Top-Ranked Features*
The feature selection technique chooses the features that have the greatest impact in predicting the output. It aids in the reduction of overfitting, typically improves accuracy, and greatly decreases training time. Univariate selection, principal component analysis (PCA), recursive feature elimination (RFE), bagged decision trees (e.g., random forest) and boosted trees (e.g., Extreme Gradient Boosting) are some of the feature selection methods used in the literature. Because of its capacity to handle datasets with many predictor variables, random forest frequently gives superior performance [52]. As a result, out of 25 variables, including age, gender, sign and symptoms, comorbidity, and clinical biomarkers, a random forest-based feature selection technique was employed in this study to rank the features in risk prediction.

**Experiments**
As mentioned earlier, two different types of investigations were carried out: risk classification and the outcome prediction for the high-risk patient. Five-fold cross-validation was performed in this study. Therefore, 80 % of the data was used for training and 20 % for testing in each fold. Finally, a weighted average of the five folds was calculated. The number of trainings, test Chest X-ray images, and clinical data used in the two experiments are listed in Table 2.

**Table 2.** Details of the dataset used for training, validation, and testing.

| Database | Types | #No patients | Train data/fold | Augmented train data/fold | Test data/fold |
|---|---|---|---|---|---|
| Risk Classification | Low | 396 | 317 | 317×4=1268 | 79 |
|  | High | 534 | 427 | 427×3=1281 | 107 |
| Outcome Prediction | Survived | 364 | 291 | 291×4=1164 | 73 |
|  | Death | 170 | 136 | 136×9=1224 | 34 |

All of the experiments in this study were conducted using the PyTorch library and Python 3.7 on an Intel® Xeon® CPU E5-2697 v4 running at 2.30GHz and the computer has 64 GB RAM and a 16 GB NVIDIA GeForce GTX 1080 GPU.

*Development and Internal Validation of Stacking Classification Model*
Eight machine learning models such as Random Forest [53], Support Vector Machine (SVM) [54], K-nearest neighbor (KNN) [55], Adaboost [56], XGBoost [56], Gradient boosting, linear discriminant analysis (LDA) [57], and Logistic regression [58] were used to reduce features after PCA from CXR images and top-ranked clinical features, individually and in combination, for risk and death prediction. The three best-performing models were chosen as base learner models ($M_1$, $M_2$, $M_3$) to develop the stacking model, and logistic regression classifier was then utilized in the second phase for training the meta learner model ($M_l$), resulting in separate performance matrices based on the final prediction.

*Experiment-01: Risk stratification using CXR Image and Clinical Data*
In this experiment, we investigated three different experiments to predict the risk of COVID-19 patients. The first one is conducted on CXR image features, while the second one is carried on Clinical features, and finally, the combined features from both modalities are used to stratify the risk.
*A. Binary Classification (Low vs High Risk) using CXR Images*
The ChexNet model was used to extract features from CXR and then PCA was used to reduce the dimensionality of the CXR features. Then, using reduced feature components and five-fold cross-validation, eight alternative ML classifiers were developed to determine which models performed well in classifying low and high-risk patients. The stacking model was built using the top three base models and a meta-model and the performance of the stacking technique for the CXR image alone is reported.
*B. Binary Classification (Low vs High Risk) using Clinical Data*

Using five-fold cross-validation, Top-5 features (LDH, O2 percentage, Age, WBC, and CRP) identified in the previous stage were tested on eight different ML classifiers to determine which models performed best in classifying low and high-risk patients. A stacking model was trained using the top-performing three algorithms as base models to train a meta learner and the performance of the meta learner and base models are reported.

*C. Binary Classification (Low vs High Risk) using CXR Images and Clinical Data*
It was important to investigate how well reduced CXR feature components and top-ranked clinical features performed in classifying low and high-risk patients using different ML classifiers for five-fold cross-validation. This experiment will reflect the strength of the multimodal approach proposed in this study compared to hundreds of approaches published on CXR alone or tens of approaches published on clinical data alone.

*Experiment-02: Death Probability Prediction for High-risk Patients*
We studied three investigations to predict the death outcome of high-risk COVID-19 patients, as shown in Experiment-01. The first one is conducted on CXR image features, while the second one is carried on Clinical features, and finally, the combined features from both modalities are used to stratify the dead and survived patients.

*A. Binary Classification (Survival vs Death) using CXR Images*
The features extracted from the CXR images using ChexNet were dimensionality reduced using PCA and used to train eight different ML classifiers to see which models performed well in predicting the mortality outcome of high-risk patients using 5-fold cross-validation. Among the eight models, the best performing three models were used to train the stacking model and the results of base and stacking models are reported.

*B. Binary Classification (Survival vs Death) using Clinical Data*
Top-5 clinical features (LDH, O2 percentage, Age, WBC, and CRP) were tested on eight different ML classifiers to determine which models performed best in predicting the mortality outcome among high-risk patients. A stacking model was trained using the top-performing three algorithms as a base model to train a meta learner and the performance of the meta learner and base models are reported.

*C. Binary Classification (Survival vs Death) using CXR Images and Clinical Data*
As a multimodal approach, we have investigated the efficacy of reduced CXR features and top-ranked clinical features to predict the mortality outcome of high-risk patients using five-fold cross-validation using same eight models. Then the Top-3 best performing models were used to train the Stacking ML model and the results for base models and stacking model were reported.

*Development and Validation of Logistic Regression-based Nomogram*
Nomograms are a popular graphical scoring technique for comprehending statistical models into a single event probability estimate [59]. This can be developed using different ML classifier, e.g., Logistic regression classifier. Logistic regression uses multiple independent predictors (x) to predict outcome (y), which are made linearly related to outcome. Event probability (P) can be calculated using linear prediction and results can be reported. A logistic regression-based nomogram was developed for high-risk patients to stratify their outcomes of survival and death. Using the integrated features from CXR and clinical data and the base learners' prediction, a nomogram using logistic regression was constructed. Furthermore, calibration curves for model construction and validation were plotted to compare the outcomes of the projected and actual probability of death for high-risk patients. Furthermore, we used decision curve analysis to finalize the ranges of threshold probabilities within the clinically useful range of the nomograms.

**Performance Metrics**
The area under the curve (AUC) from the receiver operating characteristic (ROC) along with Sensitivity, Precision, Accuracy, Specificity, and F1-Score were used to assess the performance of different classifiers. Because this study used five-fold cross-validation, the results are based on the complete dataset (five test fold-concatenated). Because different classes had variable numbers of instances, weighted metrics per class and overall accuracy were reported. As a metric for evaluation, the area under the curve (AUC) was considered. The mathematical expressions of five evaluation metrics are shown in Equations (2-6):

$$Accuracy_{class_i}(A) = \frac{TP_{class\_i} + TN_{class\_i}}{TP_{class\_i} + TN_{class\_i} + FP_{class\_i} + FN_{class\_i}} \quad (2)$$

$$Precision_{class_i}(P) = \frac{TP_{class\_i}}{TP_{class\_i} + FP_{class\_i}} \quad (3)$$

$$Recall/Sensitivity_{class_i}(R) = \frac{TP_{class_i}}{TP_{class_i} + FN_{class_i}} \quad (4)$$

$$F1\_score_{class_i}(F1) = 2\frac{Precision_{class_i} \times Sensitivity_{class_i}}{Precision_{class_i} + Sensitivity_{class_i}} \quad (5)$$

$$Specificity_{class_i}(S) = \frac{TN_{class\_i}}{TN_{class\_i} + FP_{class\_i}} \quad (6)$$

*where $class_i$ = Mild and Severe or Survived and Death*

Where TP is true positive meaning correcting detecting the actual class, TN is true negative meaning misclassifying the actual class, FP is false positive meaning correcting detecting the other class and FN is false negative meaning misclassifying the other class.

## III. Results
**Best Features and Their Combination Selection**

Thirty-four statistically significant features used to select the top ranked 10 features using the random forest feature ranking technique (Figure 5). Table 3 illustrates the results of using multiple classifiers to test these top-ranked 10 features to determine the best-performing feature combinations. The Gradient Boosting classifier surpasses different networks in binary classification (low- vs. high-risks) when employing the top-ranked 5 features. Using only the Top-5 characteristics, Gradient Boosting yields overall accuracy, weighted sensitivity, precision, specificity, and F1-score of 82.91 percent, 82.91 percent, 82.87 percent, 82.91 percent, and 82.87 percent, respectively (LDH, O2 percentage, WBC, Age, and CRP). Among the Top-10 features, it was also very important to assess the most suitable parameters for the early prediction of high-risk COVID-19 patients.

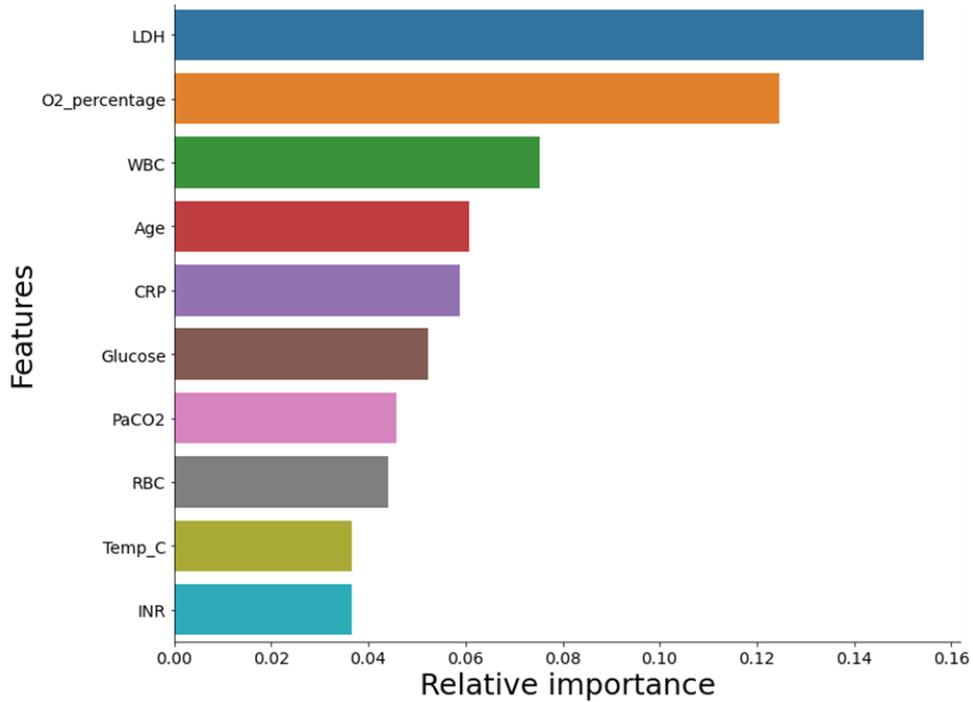

**Figure 5:** Top ten features selected using the random forest feature selection technique.

**Table 3:** Summary of the performance metrics for Top 1 to 10 clinical features

|  | Weighted Average (95% confidence interval) | | | | |
| --- | --- | --- | --- | --- | --- |
|  | P | R | F1 | S | A |
| **Top 1 feature** | 59.62 ± 6.33 | 59.71 ± 6.33 | 59.62 ± 6.33 | 59.62 ± 6.33 | 59.64 ± 6.33 |
| **Top 2 features** | 60.26 ± 6.31 | 60.26 ± 6.31 | 60.26 ± 6.31 | 60.26 ± 6.31 | 60.26 ± 6.31 |
| **Top 3 features** | 80.77 ± 5.08 | 80.8 ± 5.08 | 80.77 ± 5.08 | 80.77 ± 5.08 | 80.73 ± 5.09 |
| **Top 4 features** | 78.21 ± 5.32 | 78.22 ± 5.32 | 78.21 ± 5.32 | 78.21 ± 5.32 | 78.16 ± 5.33 |
| **Top 5 features** | **82.91 ± 1.83** | **82.87 ± 1.83** | **82.91 ± 1.83** | **82.91 ± 1.83** | **82.74 ± 1.84** |
| **Top 6 features** | 79.49 ± 5.21 | 79.58 ± 5.2 | 79.49 ± 5.21 | 79.49 ± 5.21 | 79.42 ± 5.21 |
| **Top 7 features** | 76.28 ± 5.49 | 76.39 ± 5.48 | 76.28 ± 5.49 | 76.28 ± 5.49 | 76.18 ± 5.49 |
| **Top 8 features** | 76.92 ± 5.43 | 77.22 ± 5.41 | 76.92 ± 5.43 | 76.92 ± 5.43 | 76.75 ± 5.45 |
| **Top 9 features** | 75 ± 5.58 | 75.33 ± 5.56 | 75.00 ± 5.58 | 75.00 ± 5.58 | 74.78 ± 5.6 |
| **Top 10 features** | 76.28 ± 5.49 | 76.39 ± 5.48 | 76.28 ± 5.49 | 76.28 ± 5.49 | 76.28 ± 5.49 |

**Risk Prediction of COVID-19 Patients**

In this section, the results of three different experiments to predict low or high-risk COVID-19 patients were reported. The performance of different ML models for CXR images, then using clinical data were reported separately and in combination. Each of these results is based on five-fold cross-validation.

*Performance Analysis using CXR Images*

The gradient boosting classifier was the best performing classifier for stratifying the low- and high-risk COVID-19 patients. It achieves precision, sensitivity, and F1 scores of 78.41 %, 78.48 %, and 78.41 %, respectively. The stacking model was built using the top three classifiers such as Random Forest, KNN, and Gradient Boosting. The stacking model produces slightly better performance with precision, sensitivity, and F1 scores of 79.5 %, 79.53 %, and 79.54 %, respectively.

*Performance Analysis using Clinical Data*

The gradient boosting classifier outperforms other classifiers in binary classification with precision, sensitivity, and F1 scores of 82.81 %, 82.8 %, and 82.81 %, respectively. The stacking model was trained using the top three algorithms (Random Forest, Gradient Boosting, and XGBoost). A meta learner logistic regression classifier was used and outperformed the base model with precision, sensitivity, and F1 scores of 83.01 %, 83.87 %, and 83.01 %, respectively.

*Performance Analysis using both CXR images and Clinical Data*

On combined CXR features and clinical data, the gradient boosting classifier outperforms other classifiers with precision, sensitivity, and F1 scores of 88.81 %, 88.81 %, and 88.81 %, respectively. The stacking model was built using the top three algorithms (Gradient Boosting, LDA, and Random Forest) and it outperforms the base models and produces precision, sensitivity, and F1 scores of 89.03 %, 90.44 %, and 89.03 %, respectively. The stacking model demonstrated roughly a 6% improvement using combined CXR features and top-ranked clinical features. The comparison between different classifiers using different metrics with a 95% confidence interval in the prediction of low or high-risk patients using CXR features and clinical data separately and in combination are shown in Table 4.

**Table 4:** Comparison of performance metrics for risk prediction using different ML models and approaches (single mode and multimode)

| Dataset |  | Overall | Weighted with 95% CI | | | |
| --- | --- | --- | --- | --- | --- | --- |
|  | **Classifier** | A | P | R | F1 | S |
|  | Linear Discriminant Analysis (LDA) | 71.75 ± 2.97 | 71.75 ± 2.97 | 71.75 ± 2.97 | 71.75 ± 2.97 | 71.54 ± 3 |
|  | XGBoost (XGB) | 73.69 ± 2.62 | 73.7 ± 2.62 | 73.69 ± 2.62 | 73.69 ± 2.62 | 73.68 ± 2.63 |
|  | Random Forest (RF) | 76.26 ± 3.05 | 76.3 ± 3.04 | 76.26 ± 3.05 | 76.25 ± 3.05 | 75.79 ± 3.12 |
| CXR Images | Logistic Regression (LR) | 71.18 ± 2.36 | 71.63 ± 2.4 | 71.24 ± 2.19 | 71.92 ± 2.3 | 71.31 ± 2.57 |
| | Support Vector Machine (SVM) | 74.26 ± 3.05 | 74.26 ± 3.05 | 74.26 ± 3.05 | 74.26 ± 3.05 | 74.96 ± 3.09 |
| | Extra Tree (ET) | 70.29 ± 3.19 | 70.32 ± 3.19 | 70.29 ± 3.19 | 70.3 ± 3.19 | 70.32 ± 3.19 |
| | K-Nearest Neighbors (KNN) | 75.66 ± 2.43 | 75.66 ± 2.43 | 75.66 ± 2.43 | 75.66 ± 2.43 | 74.5 ± 2.46 |
| | Gradient Boosting (GB) | 78.41 ± 3.26 | 78.48 ± 3.27 | 78.41 ± 3.26 | 78.44 ± 3.27 | 77.9 ± 3.3 |
| | **Stacking model (RF+KNN+GB)** | **79.5 ± 1.97** | **79.53 ± 1.98** | **79.54 ± 1.97** | **79.54 ± 1.97** | **79.45 ± 1.98** |
| Clinical Data | Linear Discriminant Analysis (LDA) | 78.75 ± 2.89 | 78.75 ± 2.89 | 78.75 ± 2.89 | 78.75 ± 2.89 | 78.54 ± 2.9 |
| | XGBoost (XGB) | 80.69 ± 2.79 | 80.7 ± 2.79 | 80.69 ± 2.79 | 80.69 ± 2.79 | 80.68 ± 2.79 |
| | Random Forest (RF) | 81.66 ± 2.73 | 81.66 ± 2.73 | 81.66 ± 2.73 | 81.66 ± 2.73 | 81.5 ± 2.74 |
| | Logistic Regression (LR) | 74.81 ± 2.72 | 74.8 ± 2.73 | 74.81 ± 2.72 | 74.8 ± 2.73 | 74.56 ± 2.74 |
| | Support Vector Machine (SVM) | 78.26 ± 2.91 | 78.26 ± 2.91 | 78.26 ± 2.91 | 78.26 ± 2.91 | 77.96 ± 2.93 |
| | Extra Tree (ET) | 77.29 ± 2.96 | 77.32 ± 2.96 | 77.29 ± 2.96 | 77.3 ± 2.96 | 77.32 ± 2.96 |
| | K-Nearest Neighbors (KNN) | 76.26 ± 2.64 | 76.26 ± 2.64 | 76.26 ± 2.64 | 76.26 ± 2.64 | 76.96 ± 2.66 |
| | Gradient Boosting (GB) | 82.91 ± 1.83 | 82.87 ± 1.83 | 82.91 ± 1.83 | 82.91 ± 1.83 | 82.74 ± 1.84 |
| | **Stacking model (GB+RF+XGB)** | **83.01 ± 3.15** | **83.87 ± 3.14** | **83.01 ± 3.14** | **83.01 ± 3.14** | **83.04 ± 3.17** |
| Both CXR images & Clinical data | Linear Discriminant Analysis (LDA) | 83.26 ± 2.51 | 83.26 ± 2.51 | 83.26 ± 2.51 | 83.26 ± 2.51 | 82.96 ± 2.53 |
| | XGBoost (XGB) | 85.69 ± 2.35 | 85.7 ± 2.35 | 85.69 ± 2.35 | 85.69 ± 2.35 | 85.68 ± 2.35 |
| | Random Forest (RF) | 83.75 ± 2.48 | 83.75 ± 2.48 | 83.75 ± 2.48 | 83.75 ± 2.48 | 83.54 ± 2.49 |
| | Logistic Regression (LR) | 82.29 ± 2.57 | 82.39 ± 2.56 | 82.29 ± 2.57 | 82.31 ± 2.57 | 82.49 ± 2.55 |
| | Support Vector Machine (SVM) | 81.66 ± 2.4 | 81.66 ± 2.4 | 81.66 ± 2.4 | 81.66 ± 2.4 | 81.5 ± 2.41 |
| | AdaBoost | 80.75 ± 2.61 | 80.75 ± 2.61 | 80.75 ± 2.61 | 80.75 ± 2.61 | 80.54 ± 2.62 |
| | K-Nearest Neighbors (KNN) | 76.66 ± 2.29 | 76.66 ± 2.29 | 76.66 ± 2.29 | 76.66 ± 2.29 | 76.5 ± 2.3 |
| | Gradient Boosting (GB) | 88.81 ± 2.59 | 88.8 ± 2.59 | 88.81 ± 2.59 | 88.8 ± 2.59 | 88.56 ± 2.61 |
| | **Stacking model (GB+XGB+RF)** | **89.03 ± 2.18** | **90.44 ± 2.15** | **89.03 ± 2.15** | **89.03 ± 2.15** | **88.7 ± 2.2** |

In Figure 6, it can be seen that combined CXR image features and clinical top-ranked features outperformed individual modality with an AUC of 91.5%. The AUC values for CXR image features and clinical top-ranked features individually using the stacking model produced 82.3% and 85% of AUC, respectively.

**Death Probability Prediction for High-risk Patients**
In this section, the results of three different experiments to predict the probability of death among the high-risk COVID-19 patients were reported. The five-fold performance of different ML models for CXR images, then using clinical data were reported separately.

*Performance Analysis with CXR Images*
Random Forest classifier outperforms the other 7 classifiers in classifying the dead and survived COVID-19 patients with precision, sensitivity, and F1 scores of 84.83 %, 85.02 %, and 84.83 %, respectively. The stacking model was built using the top three methods (Random Forest, Extra Tree, and Gradient Boosting) and produces precision, sensitivity, and F1 scores of 86.35 %, 83.22 %, and 86.35 %, respectively.

*Performance Analysis with Clinical Data*
The gradient boosting model outperforms the other seven classifiers in stratifying the survival and dead patients with precision, sensitivity, and F1 scores of 89.14 %, 89.86 %, and 89.14 %, respectively. The stacking model was trained using the top three models (Random Forest, XGBoost, and Extra Tree). The stacking model beat previous base models, achieving 91.2 % precision, 91.25 % sensitivity, and 91.2 % F1 scores, respectively.

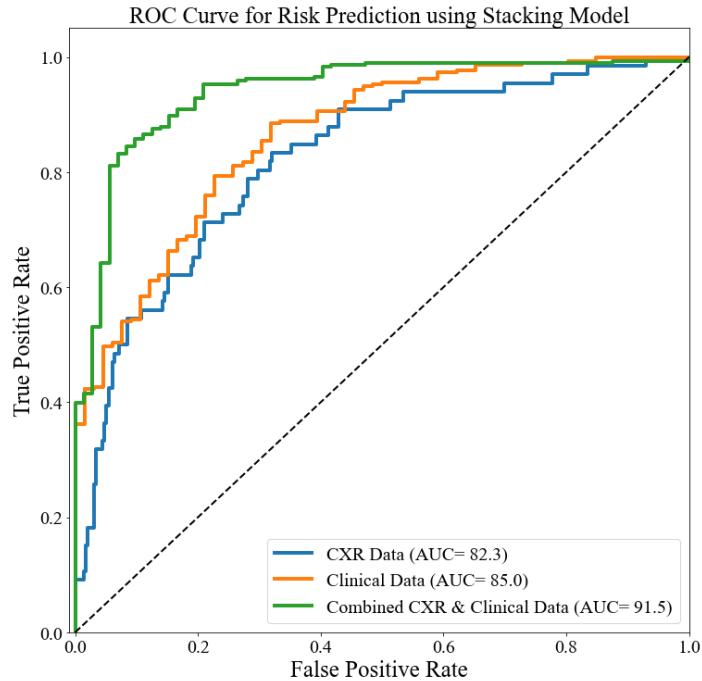

**Figure 6:** ROC curves for risk prediction of COVID-19 patients with single and multi-modal data using the stacking ML model.

*Performance Analysis using both CXR Images and Clinical Data*
Random Forest classifier outperforms other models with precision, sensitivity, and F1 scores of 91.76 %, 91.86 %, and 91.76 %, respectively. The stacking machine learning model was trained using Random Forest, Extra Tree, and Gradient Boosting and it outperforms the base model with precision, sensitivity, and F1 scores of 92.88 %, 93.37 %, and 92.88 %, respectively. In terms of all the different performance matrics, the performance of the stacking model improved by ~ 6% when using both reduced CXR features and clinical top features, refer Table 5.

**Table 5:** Comparison of performance metrics for death prediction using different ML models and approaches (single mode and multimode)

| Dataset | Classifier | Overall | Weighted with 95% CI | | | |
|---|---|---|---|---|---|---|
| | | A | P | R | F1 | S |
| | Linear Discriminant Analysis (LDA) | 68.35 ± 4.41 | 73.22 ± 4.2 | 68.35 ± 4.41 | 68.35 ± 4.41 | 69.4 ± 4.37 |
| | XGBoost (XGB) | 73.03 ± 4.21 | 75.9 ± 4.06 | 73.03 ± 4.21 | 73.03 ± 4.21 | 73.79 ± 4.17 |
| | Random Forest (RF) | 84.83 ± 3.4 | 85.02 ± 3.38 | 84.83 ± 3.4 | 84.83 ± 3.4 | 84.91 ± 3.4 |
| CXR Images | Logistic Regression (LR) | 67.6 ± 4.44 | 72.65 ± 4.23 | 67.6 ± 4.44 | 67.6 ± 4.44 | 68.69 ± 4.4 |
| | Support Vector Machine (SVM) | 54.87 ± 4.72 | 59.08 ± 4.66 | 54.87 ± 4.72 | 54.87 ± 4.72 | 56.26 ± 4.71 |
| | Extra Tree (ET) | 82.4 ± 3.61 | 82.72 ± 3.59 | 82.4 ± 3.61 | 82.4 ± 3.61 | 82.53 ± 3.6 |
| | K-Nearest Neighbors (KNN) | 72.85 ± 4.22 | 76.51 ± 4.02 | 72.85 ± 4.22 | 72.85 ± 4.22 | 73.68 ± 4.18 |
| | Gradient Boosting (GB) | 80.71 ± 3.74 | 82.59 ± 3.6 | 80.71 ± 3.74 | 80.71 ± 3.74 | 81.17 ± 3.71 |
| | **Stacking model (RF+ET+GB)** | **86.35 ± 3.26** | **83.22 ± 3.54** | **86.35 ± 3.26** | **86.35 ± 3.26** | **87.4 ± 3.15** |
| Clinical Data | Linear Discriminant Analysis (LDA) | 69.29 ± 4.38 | 73.74 ± 4.17 | 69.29 ± 4.38 | 69.29 ± 4.38 | 70.29 ± 4.33 |
| | XGBoost (XGB) | 87.08 ± 3.18 | 87.73 ± 3.11 | 87.08 ± 3.18 | 87.08 ± 3.18 | 87.26 ± 3.16 |
| | Random Forest (RF) | 89.14 ± 2.95 | 89.86 ± 2.86 | 89.14 ± 2.95 | 89.14 ± 2.95 | 89.31 ± 2.93 |
| | Logistic Regression (LR) | 68.91 ± 4.39 | 73.37 ± 4.19 | 68.91 ± 4.39 | 68.91 ± 4.39 | 69.92 ± 4.35 |
| | Support Vector Machine (SVM) | 53.56 ± 4.73 | 58.01 ± 4.68 | 53.56 ± 4.73 | 53.56 ± 4.73 | 55.02 ± 4.72 |
| | Extra Tree (ET) | 86.33 ± 3.26 | 86.31 ± 3.26 | 86.33 ± 3.26 | 86.33 ± 3.26 | 86.32 ± 3.26 |
| | K-Nearest Neighbors (KNN) | 75.84 ± 4.06 | 79.56 ± 3.82 | 75.84 ± 4.06 | 75.84 ± 4.06 | 76.6 ± 4.02 |
| | Gradient Boosting (GB) | 71.72 ± 4.27 | 71.42 ± 4.29 | 71.72 ± 4.27 | 71.72 ± 4.27 | 71.56 ± 4.28 |
| | **Stacking model (RF+XGB+ET)** | **91.2 ± 2.69** | **91.25 ± 2.68** | **91.2 ± 2.69** | **91.2 ± 2.69** | **91.22 ± 2.68** |
| Both CXR images & Clinical data | Linear Discriminant Analysis (LDA) | 74.34 ± 4.14 | 78.91 ± 3.87 | 74.34 ± 4.14 | 74.34 ± 4.14 | 75.19 ± 4.1 |
| | XGBoost (XGB) | 77.53 ± 3.96 | 80.24 ± 3.78 | 77.53 ± 3.96 | 77.53 ± 3.96 | 78.15 ± 3.92 |
| | Random Forest (RF) | 91.76 ± 2.61 | 91.86 ± 2.59 | 91.76 ± 2.61 | 91.76 ± 2.61 | 91.8 ± 2.6 |
| | Logistic Regression (LR) | 79.78 ± 3.81 | 85.84 ± 3.31 | 79.78 ± 3.81 | 79.78 ± 3.81 | 80.47 ± 3.76 |
| | Support Vector Machine (SVM) | 69.29 ± 4.38 | 75.98 ± 4.05 | 69.29 ± 4.38 | 69.29 ± 4.38 | 70.34 ± 4.33 |
| | Extra Tree (ET) | 90.07 ± 2.84 | 90.65 ± 2.76 | 90.07 ± 2.84 | 90.07 ± 2.84 | 90.22 ± 2.82 |
| | K-Nearest Neighbors (KNN) | 81.84 ± 3.66 | 83.8 ± 3.49 | 81.84 ± 3.66 | 81.84 ± 3.66 | 82.28 ± 3.62 |
| | Gradient Boosting (GB) | 88.95 ± 2.97 | 90.21 ± 2.82 | 88.95 ± 2.97 | 88.95 ± 2.97 | 89.19 ± 2.95 |
| | **Stacking model (RF+ET+GB)** | **92.88 ± 2.44** | **93.37 ± 2.36** | **92.88 ± 2.44** | **92.88 ± 2.44** | **92.65 ± 2.48** |

In Figure 7, it also can be visible that combined CXR image features and clinical top-ranked features outperformed individual modalities with an AUC of 92.8%. The reduced CXR image features and clinical top-ranked features using the stacking model individually produce an AUC of 88.4% and 91.1%, respectively.

*Stacking ML Based Nomogram*

Because the Logistic regression meta learner performs best in the classification of survival and death patients, a Nomogram that leverages the probability scores of the three best models (Random Forest (M1), Extra Tree (M2), and Gradient Boosting (M3)) was created to accurately estimate the survival and death probabilities of the high-risk group. The link between the probability scores of these base learner models and the likelihood of death in high-risk patients was investigated using multivariate logistic regression analysis (Table 6). A common method for identifying the significant features is to use the z-value, which is calculated using regression coefficient and standard error. The independent variable becomes significant when the z-value is high.

**Table 6:** Summary of logistic regression analysis

| Outcome | Coefficient | Bootstrap Std. Error | Z | P>|z| | [95% CI] | |
|---|---|---|---|---|---|---|
| **Random Forest** | -14.85299 | 3.262317 | -4.5 | 0.000 | -21.24701 | -8.458965 |
| **Extra Tree** | -5.028269 | 5.40965 | -0.93 | 0.353 | -15.63099 | 5.57445 |
| **Gradient Boosting** | -1.788734 | 0.47932 | -3.73 | 0.000 | -2.728183 | -0.84928 |
| **cons** | 11.23907 | 1.822279 | 6.17 | 0.0004 | 7.667468 | 14.81067 |

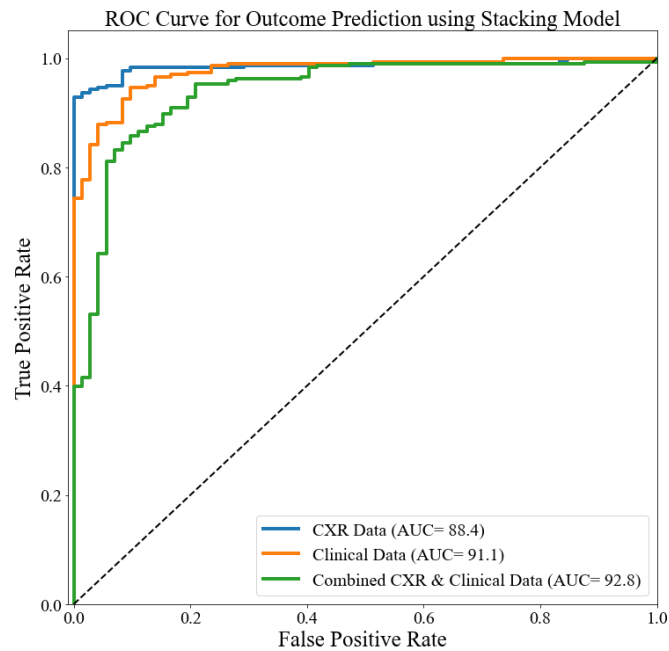

**Figure 7:** ROC curves for outcome prediction of high-risk patients with single and multi-modal data using the stacking ML model.

It can be seen from Table 6 that Extra Tree (M2) is not a very good predictor of COVID-19 individuals out of three probability scores, although Random Forest (M1) and Gradient Boosting (M3) are good predictors. The P-value can be used to determine significant variable if p < 0.05, X-variables can have a significant connection with Y-variables. It is evident also from the p-value that the Extra Tree model is not a strong predictor.
However, it was observed that the model performance was slightly reduced by stacking two models instead of three. Therefore, the Nomogram is created using three models. The nomogram has 6 rows, spanning from 1 to 3, reflecting the incorporated variables, as shown in Figure 7. "Points axis" yielded a score for each variable in the death or survived high-risk group. The scores were calculated by adding the points from the three factors (row 4), and the final score was displayed in row 6. To calculate the chance of a patient dying, a line is drawn from the "Total Score" axis to the "Prob axis" (row 5).
Alternatively, the following formula can be used to calculate the nomogram score:

**Linear prediction** = 11.23907 -14.85299 × Random Forest (M1) -5.028269 × Extra Tree (M2) -1.788734 × Gradient Boosting (M3)     (11)

**Probability of death in high-risk patients** = 1/ (1+exp (-Linear Prediction))     (12)

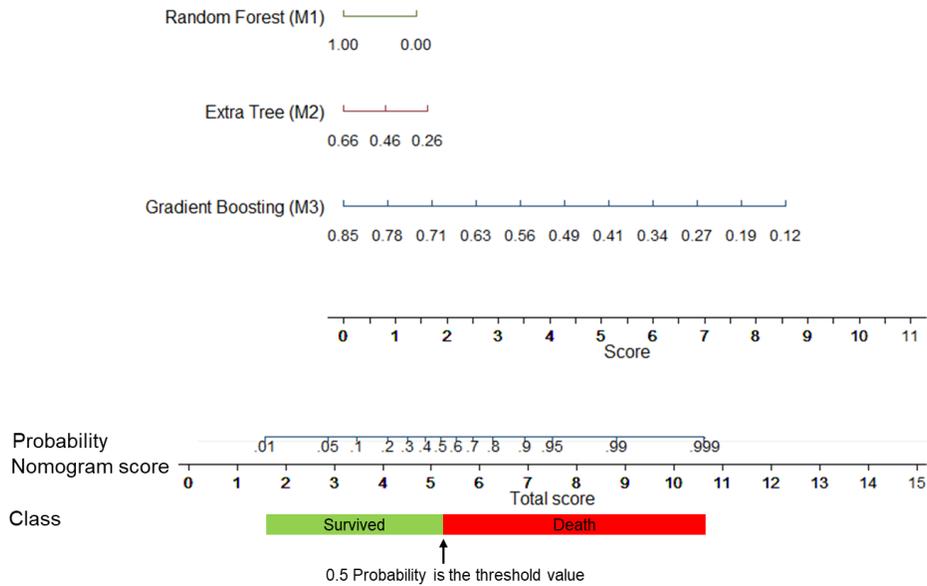

**Figure 7:** A Nomogram for prediction of death in COVID-19 severe patients was created using Random Forest (M1), ExtraTree (M2), and Gradient boosting (M3).

Figure 7 also shows the Nomogram scores for both survived and death classes. It was found that 50 % cutoffs of classification probability represent a Nomogram score of 4.8 or probability of 0.5, which stratifies the classes.

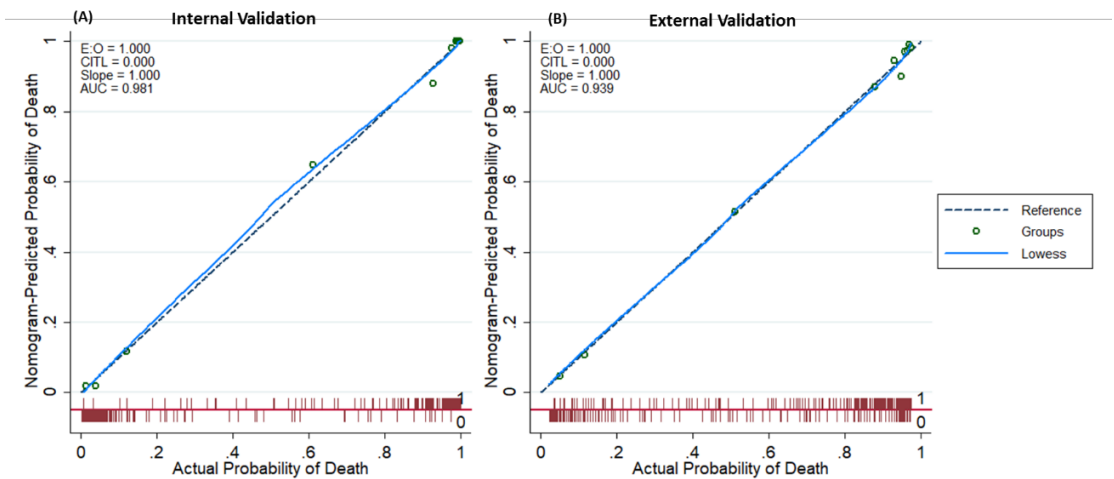

**Figure 8:** (A) Internal validation calibration plot, (B) External validation calibration plot.

Figure 8 shows the calibration plot for both internal and external validation. It is evident from Figure 8 that each calibration curve is very close to the diagonal line reflecting a reliable model. The AUC values for internal and external validation are 98.1% and 93.8 %, respectively, which is also reflecting the superior performance of the proposed model.

Figure 9 demonstrates that each predictor model's net benefit was positive (threshold<0.95), indicating that each predictor contributed to the outcome prediction. The entire model, in particular, provided the best results, necessitating the use of three base models as predictors in the Stacking model.

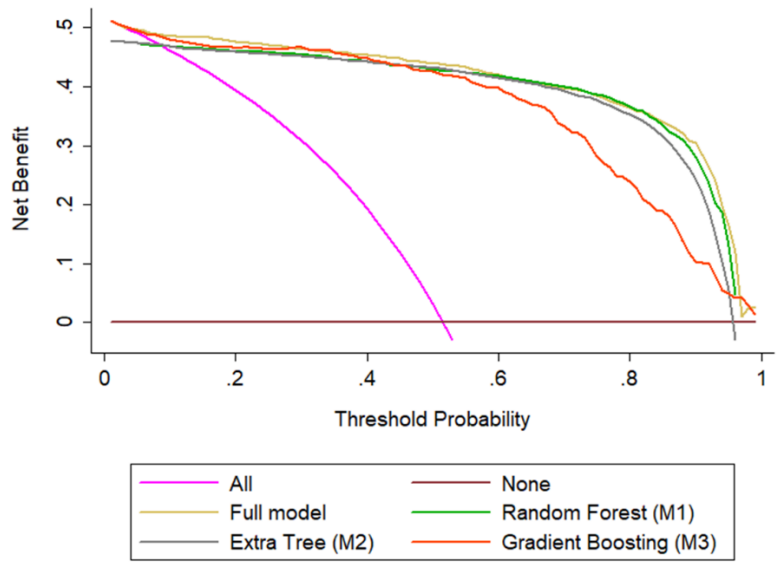

**Figure 9:** Decision curves analysis comparing different models to predict the death probability of patients with high-risk COVID-19.

*Performance Evaluation of the Model*
We compared the actual death to the expected death among high-risk individuals using the Nomogram score. The proportions of death outcomes in the training set were 91.9 percent (125/136) for the death group and 8.1 percent (11/136) for the survived group, as shown in Table 7(A), whereas the proportions of death outcomes in the test set were 91.18 percent (31/34) for the death group and 8.82 percent (3/34) for the survived group (Table 7(B)). Actual death rates were significantly different between the two groups ($p<0.001$). As a result, this scoring technique can be useful to forecast the patient outcomes.

**Table 7:** Performance evaluation of the model in the training cohort (A) and testing cohort (B) using Fisher's exact probability test

(A)

| Prediction | Outcome | |
|---|---|---|
| | Survived | Death |
| Survived | 280 (96.22%) | 11 (8.1%) |
| Death | 11 (3.78%) | 125 (91.9%) |
| Overall | 291(100%) | 136 (100%) |

(B)

| Prediction | Outcome | |
|---|---|---|
| | Survived | Death |
| Survived | 68 (93.1%) | 3 (8.82%) |
| Death | 5 (6.9%) | 31 (91.18%) |
| Overall | 73 (100%) | 34 (100%) |

*Web Application with Back-end Server*
As an extnesion to this work, we developed an online application (https://qu-mlg.com/projects/covid-severity-grading-AI) that allows clinicians to input demographic and clinical data (LDH, O2 percentage, WBC, age, and CRP) as well as CXR images. BIO-CXRNET is a Google Cloud-based AI application that analyzes data to determine whether a user is a low-risk or high-risk patient. Our model identifies the patient's death risk probability if the patient is in the extreme risk group.

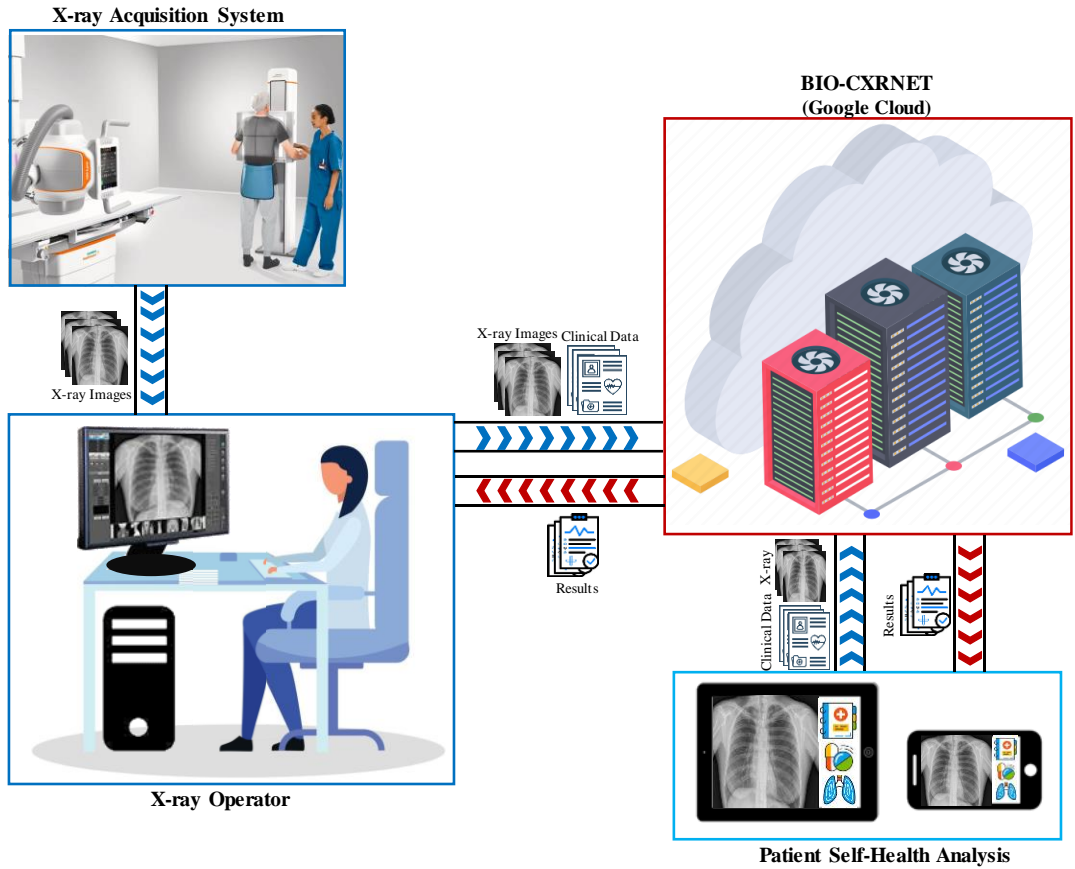

Figure 10: COVID-19 severity risk detection tool using web application framework.

The backend application is developed in python using Flask. Flask is a robust backend application framework for python. The cloud application is deployed to an Apache 2.0 HTTP server running on an Ubuntu 20.01 LTS Google Computation Engine (GCE). To optimize server cost a minimum configuration GCE instance is hired. The GCE server have a 4 core Intel Xenon processor unit with 8 GB DDR4 memory and a 100GB balanced persistent storage. To accommodate the computation intensive ML models in such a resource constraint environment, kernel level configurations are modified for the operating system. Such configurations include enabling non-threaded pre-forking for the Apache webserver, to facilitate additional memory for the Tensorflow processes. This web-application was created with Flutter which is based on Google's Dart programming language.

In the prototype system, the radiologists/clinicians/users will first enter demographic information, then system will ask the user to upload CXR images as well as four biomarkers: LDH, $O_2$ percentage, WBC, and CRP. This will be uploaded to the server, where it will be pre-processed and applied to BIO-CXRNET model to determine if the user is a low or high-risk patient (Figure 10). The data will be analyzed by the AI backend, and an answer will be displayed on the screen. The application will both display and save the findings in an SQLite-based local database. In summary, the application can assist in promptly assessing the severity risk of COVID patients using a limited number of blood indicators, hence reducing the burden on the healthcare system.

## IV. Discussion

This work presents a multimodal system for predicting COVID-19 positive patients' risk and consequently stratify the potential outcome of high-risk patients. The performance of both of the experiments were analyzed separately and in combination using CXR image and clinical data. It was observed from both the experiments that the multimodal approach outperforms individual modality. For risk group stratification among COVID-19 patients, CXR and clinical features combined had an accuracy of 89.03% in comparison to 80.11% and

86.01% for CXR and clinical features, respectively. Furthermore, in case of outcome prediction for the high-risk patients, multimodal technique outperformed individual modality with an accuracy of 92.3% whereas CXR images and clinical data individually produced an accuracy of 89.5% and 90.11%, respectively. The results reported in this work has superior performance compared to some of the state-of-the-art performance reported in the literature, as shown in Table 8.

Table 8: Comparison with state of the art works in the literature

| Paper | Method | Dataset | Results |
|---|---|---|---|
| Yan et al. [60] | Predict individual patient mortality using the XGBoost classifier | Clinical biomarkers (485 COVID-19 patients) | Accuracy of 90% |
| Rahman et al. [48] | Predict COVID-19 severity using machine learning classifiers | Clinical biomarkers (375 COVID-19 patients) | Accuracy of 90.8% |
| Abbas et al. [61] | CNN (DeTraC) | Chest X-ray images (749 COVID-19) | Accuracy of 93% |
| Zulfaezal et al. [62] | CNN | Chest X-ray images (1565 COVID-19) | Accuracy of 71.9% |
| Soda et al. [42] | Deep multimodal CNN | Chest X-ray images and clinical biomarkers (820 COVID-19) | Accuracy of 76.8% |
| **Proposed Study** | **CNN and ML classifiers to predict the severity, and developed a nomogram scoring tools** | **Chest X-ray images and clinical biomarkers (930 COVID-19)** | **Accuracy of 92.88%** |

In our previous research [63] on severe acute respiratory syndrome (SARS) [64], Middle East respiratory syndrome (MERS) [65], and COVID-19 [66], we found that older age was a predictor of poor outcomes in COVID-19 patients. Because LDH indicates tissue/cell death, it is a common sign of tissue/cell damage. Serum LDH has been established as a key biomarker for idiopathic pulmonary fibrosis activity and severity. According to Yan et al. [60] [], the increase in LDH is considered in patients with severe pulmonary interstitial disease and is one of the most important prognostic markers of lung injury. The increase in LDH levels in COVID-19 patients who are severely ill implies an increase in lung damage severity.

CRP testing at admission is associated with the prediction of short-term mortality related to COVID-19-related illnesses, according to research conducted by Lu et al. [67]. CRP is synthesized by hepatocytes that are activated by certain cytokines originating from activated leukocytes, such as those produced by infections, inflammations, or an injury to the tissue. CRP is synthesized by hepatocytes that are activated by certain cytokines originating from activated leukocytes, such as those produced by infections, inflammations, or an injury to the tissue. Our study found that increased CRP levels at admission were associated with increased mortality risk among patients with COVID-19. These findings indicated a severe inflammation or possibly a secondary infection had developed in these patients, and empirical antibiotic treatment might be required. The increase of CRP, an important marker for poor prognosis in acute respiratory distress syndrome reflects a persistent state of inflammation [68, 69]. COVID-19 patients have large gray-white lesions in their as the result of this persistent inflammatory response [70].

The five biomarkers identified in our investigation were associated with inflammation, immunology, and coagulation function, all of which may have a role in COVID-19 pathogenesis, based on the previous studies. We predicted that the inflammatory response to Severe acute respiratory syndrome coronavirus 2 (SARS-CoV-2) infection is fundamental to COVID-19 pathogenesis, and that dysregulation of the immune and/or coagulation systems leads to severe clinical outcomes, including Acute respiratory distress syndrome (ARDS), coagulopathy, and septic shock, among others. Patients who died showed lower WBC and O2 percentages, as well as higher age, CRP, and LDH values than survivors. COVID-19 patients with a high mortality risk may benefit from early care based on a complete assessment of the inflammatory response, immunological dysfunction, and coagulopathy. As expected clinical information along with the Chest X-ray images helps in the reliable detection of COVID-19 severity and mortality risk.

In addition, our nomogram can be used in a variety of therapeutic scenarios. It outperforms other models proposed in the literature, to the best of our knowledge. Furthermore, the nomogram's score served as a quantitative tool for identifying patients with a high mortality risk upon admission and guiding clinical therapy. COVID-19 individuals were assigned to risk categories based on their hospital admission data. Isolation and treatment of low-risk cases should be done in isolation centers. Survivors from high-risk categories should be admitted to a hospital with an isolation unit to receive complete care. The high-risk group requires close monitoring and is referred to the ICU for intense treatment and critical assistance.

## V. Conclusion

A multimodal method was proposed that used a unique machine learning architecture to predict severity and mortality risk in COVID-19 patients utilizing both CXR images and clinical data. The suggested architecture uses CXR images and only five parameters: LDH, O2 percentage, Age, WBC, and CRP, which shows outstanding results for detecting low and high-risk COVID-19 positive individuals with very high sensitivity. Furthermore, the proposed nomogram-based approach accurately forecasts the likelihood of death in individuals at high risk. Our nomogram for predicting the prognosis of COVID-19 patient's demonstrated good discrimination and calibration based on various risk indicators. Since the model uses CXR image and clinical parameters which can counteract the criticisms of the clinicians on using only radiographic images for prognostic purpose. This model can identify the potential risk of the patient at admission which can significantly help is hospital resource management. Although the study has used data from initial variants but the clinical biomarkers identified in this work are supported by a large pool of clinical studies done on other variants and therefore, we expect this model can be equally useful in Omicron and other future variants, which can emerge in the coming winter. As a result, doctors could use this technique to make a quick and fair judgment to optimize patient stratification management and possibly minimize mortality rates. However, this quantitative tool should be validated in large-scale multicenter and multi-country prospective study to demonstrate it usability further in clinical setting.

**Funding**
This work was supported by the Qatar National Research Grant: UREP28-144-3-046. The statements made herein are solely the responsibility of the authors.

**Conflict of Interest**
No conflict of interest to declare.

**Reference**
[1]     (2020, 1 October, 2021). *WHO Coronavirus Disease (COVID-19) Dashboard* [Online]. Available: https://covid19.who.int/?gclid=Cj0KCQjwtZH7BRDzARIsAGjbK2ZXWRpJROEl97HGmSOx0_ydkVbc02Ka1FlcysGjEI7hnaIeR6xWhr4aAu57EALw_wcB
[2]     T. Singhal, "A review of coronavirus disease-2019 (COVID-19)," *The indian journal of pediatrics,* vol. 87, pp. 281-286, 2020.
[3]     C. Sohrabi, Z. Alsafi, N. O'neill, M. Khan, A. Kerwan, A. Al-Jabir*, et al.*, "World Health Organization declares global emergency: A review of the 2019 novel coronavirus (COVID-19)," *International journal of surgery,* vol. 76, pp. 71-76, 2020.
[4]     P. Kakodkar, N. Kaka, and M. Baig, "A comprehensive literature review on the clinical presentation, and management of the pandemic coronavirus disease 2019 (COVID-19)," *Cureus,* vol. 12, 2020.
[5]     Y. Li, L. Yao, J. Li, L. Chen, Y. Song, Z. Cai*, et al.*, "Stability issues of RT-PCR testing of SARS-CoV-2 for hospitalized patients clinically diagnosed with COVID-19," *Journal of medical virology,* vol. 92, pp. 903-908, 2020.
[6]     A. Tahamtan and A. Ardebili, "Real-time RT-PCR in COVID-19 detection: issues affecting the results," *Expert review of molecular diagnostics,* vol. 20, pp. 453-454, 2020.
[7]     J. Xia, J. Tong, M. Liu, Y. Shen, and D. Guo, "Evaluation of coronavirus in tears and conjunctival secretions of patients with SARS-CoV-2 infection," *Journal of medical virology,* vol. 92, pp. 589-594, 2020.

[8]     T. Ai, Z. Yang, H. Hou, C. Zhan, C. Chen, W. Lv, *et al.*, "Correlation of chest CT and RT-PCR testing for coronavirus disease 2019 (COVID-19) in China: a report of 1014 cases," *Radiology,* vol. 296, pp. E32-E40, 2020.

[9]     S. Salehi, A. Abedi, S. Balakrishnan, and A. Gholamrezanezhad, "Coronavirus disease 2019 (COVID-19): a systematic review of imaging findings in 919 patients," *Ajr Am J Roentgenol,* vol. 215, pp. 87-93, 2020.

[10]    Y. Fang, H. Zhang, J. Xie, M. Lin, L. Ying, P. Pang, *et al.*, "Sensitivity of chest CT for COVID-19: comparison to RT-PCR," *Radiology,* vol. 296, pp. E115-E117, 2020.

[11]    D. J. Brenner and E. J. Hall, "Computed tomography—an increasing source of radiation exposure," *New England journal of medicine,* vol. 357, pp. 2277-2284, 2007.

[12]    F. Shi, J. Wang, J. Shi, Z. Wu, Q. Wang, Z. Tang, *et al.*, "Review of artificial intelligence techniques in imaging data acquisition, segmentation, and diagnosis for COVID-19," *IEEE reviews in biomedical engineering,* vol. 14, pp. 4-15, 2020.

[13]    C. Huang, Y. Wang, X. Li, L. Ren, J. Zhao, Y. Hu, *et al.*, "Clinical features of patients infected with 2019 novel coronavirus in Wuhan, China," *The lancet,* vol. 395, pp. 497-506, 2020.

[14]    M. Hosseiny, S. Kooraki, A. Gholamrezanezhad, S. Reddy, and L. Myers, "Radiology perspective of coronavirus disease 2019 (COVID-19): lessons from severe acute respiratory syndrome and Middle East respiratory syndrome," *Ajr Am J Roentgenol,* vol. 214, pp. 1078-1082, 2020.

[15]    P. Rajpurkar, J. Irvin, R. L. Ball, K. Zhu, B. Yang, H. Mehta, *et al.*, "Deep learning for chest radiograph diagnosis: A retrospective comparison of the CheXNeXt algorithm to practicing radiologists," *PLoS medicine,* vol. 15, p. e1002686, 2018.

[16]    J. a. R. Irvin, Pranav and Ko, Michael and Yu, Yifan and Ciurea-Ilcus, Silviana and Chute, Chris and Marklund, Henrik and Haghgoo, Behzad and Ball, Robyn and Shpanskaya, Katie and Seekins, Jayne and Mong, David A. and Halabi, Safwan S. and Sandberg, Jesse K. and Jones, Ricky and Larson, David B. and Langlotz, Curtis P. and Patel, Bhavik N. and Lungren, Matthew P. and Ng, Andrew Y, "CheXpert: A Large Chest Radiograph Dataset with Uncertainty Labels and Expert Comparison," 2019.

[17]    T. Rahman, A. Khandakar, M. A. Kadir, K. R. Islam, K. F. Islam, R. Mazhar, *et al.*, "Reliable tuberculosis detection using chest X-ray with deep learning, segmentation and visualization," *IEEE Access,* vol. 8, pp. 191586-191601, 2020.

[18]    Y. Oh, S. Park, and J. C. Ye, "Deep learning COVID-19 features on CXR using limited training data sets," *IEEE transactions on medical imaging,* vol. 39, pp. 2688-2700, 2020.

[19]    S. Rajaraman, J. Siegelman, P. O. Alderson, L. S. Folio, L. R. Folio, and S. K. Antani, "Iteratively pruned deep learning ensembles for COVID-19 detection in chest X-rays," *Ieee Access,* vol. 8, pp. 115041-115050, 2020.

[20]    S. Jaeger, A. Karargyris, S. Candemir, L. Folio, J. Siegelman, F. Callaghan, *et al.*, "Automatic tuberculosis screening using chest radiographs," *IEEE transactions on medical imaging,* vol. 33, pp. 233-245, 2013.

[21]    S. Candemir, S. Jaeger, K. Palaniappan, J. P. Musco, R. K. Singh, Z. Xue, *et al.*, "Lung segmentation in chest radiographs using anatomical atlases with nonrigid registration," *IEEE transactions on medical imaging,* vol. 33, pp. 577-590, 2013.

[22]    A. Zargari Khuzani, M. Heidari, and S. A. Shariati, "COVID-Classifier: An automated machine learning model to assist in the diagnosis of COVID-19 infection in chest x-ray images," *Scientific Reports,* vol. 11, pp. 1-6, 2021.

[23]    M. D. Li, N. T. Arun, M. Gidwani, K. Chang, F. Deng, B. P. Little, *et al.*, "Automated assessment and tracking of COVID-19 pulmonary disease severity on chest radiographs using convolutional siamese neural networks," *Radiology: Artificial Intelligence,* vol. 2, p. e200079, 2020.

[24]    C. K. Kim, J. W. Choi, Z. Jiao, D. Wang, J. Wu, T. Y. Yi, *et al.*, "An automated COVID-19 triage pipeline using artificial intelligence based on chest radiographs and clinical data," *NPJ Digital Medicine,* vol. 5, pp. 1-9, 2022.

[25]    G. Maguolo and L. Nanni, "A critic evaluation of methods for COVID-19 automatic detection from X-ray images," *Information Fusion,* vol. 76, pp. 1-7, 2021.

[26] M. Roberts, D. Driggs, M. Thorpe, J. Gilbey, M. Yeung, S. Ursprung, *et al.*, "Common pitfalls and recommendations for using machine learning to detect and prognosticate for COVID-19 using chest radiographs and CT scans," *Nature Machine Intelligence,* vol. 3, pp. 199-217, 2021.

[27] T. Rahman, M. E. Chowdhury, A. Khandakar, K. R. Islam, K. F. Islam, Z. B. Mahbub, *et al.*, "Transfer learning with deep convolutional neural network (CNN) for pneumonia detection using chest X-ray," *Applied Sciences,* vol. 10, p. 3233, 2020.

[28] M. E. Chowdhury, T. Rahman, A. Khandakar, R. Mazhar, M. A. Kadir, Z. B. Mahbub, *et al.*, "Can AI help in screening viral and COVID-19 pneumonia?," *IEEE Access,* vol. 8, pp. 132665-132676, 2020.

[29] A. M. Tahir, M. E. Chowdhury, A. Khandakar, T. Rahman, Y. Qiblawey, U. Khurshid, *et al.*, "COVID-19 infection localization and severity grading from chest X-ray images," *Computers in biology and medicine,* vol. 139, p. 105002, 2021.

[30] T. Rahman, A. Khandakar, Y. Qiblawey, A. Tahir, S. Kiranyaz, S. B. A. Kashem, *et al.*, "Exploring the effect of image enhancement techniques on COVID-19 detection using chest X-ray images," *Computers in biology and medicine,* vol. 132, p. 104319, 2021.

[31] S. Al Youha, S. A. Doi, M. H. Jamal, S. Almazeedi, M. Al Haddad, M. AlSeaidan, *et al.*, "Validation of the Kuwait Progression Indicator Score for predicting progression of severity in COVID19," *MedRxiv,* 2020.

[32] Z. Weng, Q. Chen, S. Li, H. Li, Q. Zhang, S. Lu, *et al.*, "ANDC: an early warning score to predict mortality risk for patients with coronavirus disease 2019," *Journal of translational medicine,* vol. 18, pp. 1-10, 2020.

[33] J. Xie, D. Hungerford, H. Chen, S. T. Abrams, S. Li, G. Wang, *et al.*, "Development and external validation of a prognostic multivariable model on admission for hospitalized patients with COVID-19," 2020.

[34] M. S. Satu, M. I. Khan, M. R. Rahman, K. C. Howlader, S. Roy, S. S. Roy, *et al.*, "Diseasome and comorbidities complexities of SARS-CoV-2 infection with common malignant diseases," *Briefings in Bioinformatics,* vol. 22, pp. 1415-1429, 2021.

[35] S. Uddin, T. Imam, and M. Ali Moni, "The implementation of public health and economic measures during the first wave of COVID-19 by different countries with respect to time, infection rate and death rate," in *2021 Australasian Computer Science Week Multiconference*, 2021, pp. 1-8.

[36] S. Aktar, M. M. Ahamad, M. Rashed-Al-Mahfuz, A. Azad, S. Uddin, A. Kamal, *et al.*, "Machine learning approach to predicting COVID-19 disease severity based on clinical blood test data: statistical analysis and model development," *JMIR medical informatics,* vol. 9, p. e25884, 2021.

[37] W.-j. Guan, Z.-y. Ni, Y. Hu, W.-h. Liang, C.-q. Ou, J.-x. He, *et al.*, "Clinical characteristics of coronavirus disease 2019 in China," *New England journal of medicine,* vol. 382, pp. 1708-1720, 2020.

[38] D. Wang, B. Hu, C. Hu, F. Zhu, X. Liu, J. Zhang, *et al.*, "Clinical characteristics of 138 hospitalized patients with 2019 novel coronavirus–infected pneumonia in Wuhan, China," *Jama,* vol. 323, pp. 1061-1069, 2020.

[39] M. Kermali, R. K. Khalsa, K. Pillai, Z. Ismail, and A. Harky, "The role of biomarkers in diagnosis of COVID-19–A systematic review," *Life sciences,* vol. 254, p. 117788, 2020.

[40] Z. Jiao, J. W. Choi, K. Halsey, T. M. L. Tran, B. Hsieh, D. Wang, *et al.*, "Prognostication of patients with COVID-19 using artificial intelligence based on chest x-rays and clinical data: a retrospective study," *The Lancet Digital Health,* vol. 3, pp. e286-e294, 2021.

[41] M. Chieregato, F. Frangiamore, M. Morassi, C. Baresi, S. Nici, C. Bassetti, *et al.*, "A hybrid machine learning/deep learning COVID-19 severity predictive model from CT images and clinical data," *Scientific reports,* vol. 12, pp. 1-15, 2022.

[42] P. Soda, N. C. D'Amico, J. Tessadori, G. Valbusa, V. Guarrasi, C. Bortolotto, *et al.*, "AIforCOVID: predicting the clinical outcomes in patients with COVID-19 applying AI to chest-X-rays. An Italian multicentre study," *Medical image analysis,* vol. 74, p. 102216, 2021.

[43] T.-Y. Lin, P. Dollár, R. Girshick, K. He, B. Hariharan, and S. Belongie, "Feature pyramid networks for object detection," in *Proceedings of the IEEE conference on computer vision and pattern recognition*, 2017, pp. 2117-2125.

[44]   G. Huang, Z. Liu, L. Van Der Maaten, and K. Q. Weinberger, "Densely connected convolutional networks," in *Proceedings of the IEEE conference on computer vision and pattern recognition*, 2017, pp. 4700-4708.

[45]   O. Ronneberger, P. Fischer, and T. Brox, "U-net: Convolutional networks for biomedical image segmentation," in *International Conference on Medical image computing and computer-assisted intervention*, 2015, pp. 234-241.

[46]   Z. Zhou, M. M. Rahman Siddiquee, N. Tajbakhsh, and J. Liang, "Unet++: A nested u-net architecture for medical image segmentation," in *Deep learning in medical image analysis and multimodal learning for clinical decision support*, ed: Springer, 2018, pp. 3-11.

[47]   M. E. Chowdhury, T. Rahman, A. Khandakar, S. Al-Madeed, S. M. Zughaier, H. Hassen, *et al.*, "An early warning tool for predicting mortality risk of COVID-19 patients using machine learning," *Cognitive Computation,* pp. 1-16, 2021.

[48]   T. Rahman, F. A. Al-Ishaq, F. S. Al-Mohannadi, R. S. Mubarak, M. H. Al-Hitmi, K. R. Islam, *et al.*, "Mortality Prediction Utilizing Blood Biomarkers to Predict the Severity of COVID-19 Using Machine Learning Technique," *Diagnostics,* vol. 11, p. 1582, 2021.

[49]   T. Rahman, A. Khandakar, F. F. Abir, M. A. A. Faisal, M. S. Hossain, K. K. Podder, *et al.*, "QCovSML: A reliable COVID-19 detection system using CBC biomarkers by a stacking machine learning model," *Computers in Biology and Medicine,* vol. 143, p. 105284, 2022.

[50]   T. Rahman, A. Khandakar, M. E. Hoque, N. Ibtehaz, S. B. Kashem, R. Masud, *et al.*, "Development and Validation of an Early Scoring System for Prediction of Disease Severity in COVID-19 using Complete Blood Count Parameters," *Ieee Access,* vol. 9, pp. 120422-120441, 2021.

[51]   J. R. Stevens, A. Suyundikov, and M. L. Slattery, "Accounting for missing data in clinical research," *Jama,* vol. 315, pp. 517-518, 2016.

[52]   J. L. Speiser, M. E. Miller, J. Tooze, and E. Ip, "A comparison of random forest variable selection methods for classification prediction modeling," *Expert systems with applications,* vol. 134, pp. 93-101, 2019.

[53]   M. Pal, "Random forest classifier for remote sensing classification," *International journal of remote sensing,* vol. 26, pp. 217-222, 2005.

[54]   S. S. Keerthi, S. K. Shevade, C. Bhattacharyya, and K. R. K. Murthy, "Improvements to Platt's SMO algorithm for SVM classifier design," *Neural computation,* vol. 13, pp. 637-649, 2001.

[55]   G. Guo, H. Wang, D. Bell, Y. Bi, and K. Greer, "KNN model-based approach in classification," in *OTM Confederated International Conferences" On the Move to Meaningful Internet Systems"*, 2003, pp. 986-996.

[56]   T. Chen, T. He, M. Benesty, V. Khotilovich, Y. Tang, H. Cho, *et al.*, "Xgboost: extreme gradient boosting," *R package version 0.4-2,* vol. 1, pp. 1-4, 2015.

[57]   Q. Gu, Z. Li, and J. Han, "Linear discriminant dimensionality reduction," in *Joint European conference on machine learning and knowledge discovery in databases*, 2011, pp. 549-564.

[58]   C. Subasi, "Logistic regression classifier," ed: Accessed: Jun, 2021.

[59]   A. Zlotnik and V. Abraira, "A general-purpose nomogram generator for predictive logistic regression models," *The Stata Journal,* vol. 15, pp. 537-546, 2015.

[60]   L. Yan, H.-T. Zhang, J. Goncalves, Y. Xiao, M. Wang, Y. Guo, *et al.*, "An interpretable mortality prediction model for COVID-19 patients," *Nature machine intelligence,* vol. 2, pp. 283-288, 2020.

[61]   A. Abbas, M. M. Abdelsamea, and M. M. Gaber, "Classification of COVID-19 in chest X-ray images using DeTraC deep convolutional neural network," *arXiv preprint arXiv:2003.13815,* 2020.

[62]   M. Z. Che Azemin, R. Hassan, M. I. Mohd Tamrin, and M. A. Md Ali, "COVID-19 deep learning prediction model using publicly available radiologist-adjudicated chest X-ray images as training data: preliminary findings," *International Journal of Biomedical Imaging,* vol. 2020, 2020.

[63]   A. M. Tahir, Y. Qiblawey, A. Khandakar, T. Rahman, U. Khurshid, F. Musharavati, *et al.*, "Deep learning for reliable classification of COVID-19, MERS, and SARS from chest X-ray images," *Cognitive Computation,* pp. 1-21, 2022.

[64] J. C. Chan, E. L. Tsui, V. C. Wong, and H. A. S. C. Group, "Prognostication in severe acute respiratory syndrome: A retrospective time-course analysis of 1312 laboratory-confirmed patients in Hong Kong," *Respirology,* vol. 12, pp. 531-542, 2007.

[65] A. Assiri, J. A. Al-Tawfiq, A. A. Al-Rabeeah, F. A. Al-Rabiah, S. Al-Hajjar, A. Al-Barrak*, et al.*, "Epidemiological, demographic, and clinical characteristics of 47 cases of Middle East respiratory syndrome coronavirus disease from Saudi Arabia: a descriptive study," *The Lancet infectious diseases,* vol. 13, pp. 752-761, 2013.

[66] R. Chen, W. Liang, M. Jiang, W. Guan, C. Zhan, T. Wang*, et al.*, "Risk factors of fatal outcome in hospitalized subjects with coronavirus disease 2019 from a nationwide analysis in China," *Chest,* vol. 158, pp. 97-105, 2020.

[67] J. Lu, S. Hu, R. Fan, Z. Liu, X. Yin, Q. Wang*, et al.*, "ACP risk grade: a simple mortality index for patients with confirmed or suspected severe acute respiratory syndrome coronavirus 2 disease (COVID-19) during the early stage of outbreak in Wuhan, China," 2020.

[68] J.-H. Ko, G. E. Park, J. Y. Lee, J. Y. Lee, S. Y. Cho, Y. E. Ha*, et al.*, "Predictive factors for pneumonia development and progression to respiratory failure in MERS-CoV infected patients," *Journal of Infection,* vol. 73, pp. 468-475, 2016.

[69] J. Wang, X. Wu, Y. Tian, X. Li, X. Zhao, and M. Zhang, "Dynamic changes and diagnostic and prognostic significance of serum PCT, hs-CRP and s-100 protein in central nervous system infection," *Experimental and therapeutic medicine,* vol. 16, pp. 5156-5160, 2018.

[70] B. Yildiz, H. Poyraz, N. Cetin, N. Kural, and O. Colak, "High sensitive C-reactive protein: a new marker for urinary tract infection, VUR and renal scar," *Eur Rev Med Pharmacol Sci,* vol. 17, pp. 2598-2604, 2013.